\documentclass[preprint]{revtex4}
\usepackage{graphicx}
\usepackage{subfigure}
\usepackage{multirow}
\usepackage{hyperref}
\usepackage{amsmath,epstopdf,bm,amssymb,amsfonts}
\usepackage{color}
\newcommand{\dalm}{\kern1pt\vbox{\hrule height 0.9pt\hbox{\vrule width 0.9pt
			\hskip 2.5pt\vbox{\vskip 5.5pt}\hskip 3pt\vrule width 0.3pt}\hrule height 0.3pt}
	\kern1pt}
\newcommand{\be}{\begin{equation}}
\newcommand{\ee}{\end{equation}}

\begin{document}

\title{{\bf
Ferromagnetic Neutron Stars in Scalar-Tensor Theories of Gravity }}

\author{{\bf Zeinab Rezaei $^{1,2}$ \footnote{Corresponding author. E-mail:
zrezaei@shirazu.ac.ir}} and {\bf Habib Yousefi Dezdarani $^{1}$} }
 \affiliation{ $^{1}$Department of Physics, Shiraz
University, Shiraz 71454, Iran.\\
$^{2}$Biruni Observatory, Shiraz
University, Shiraz 71454, Iran.}


\begin{abstract}
Ferromagnetic spin ordering can take place in neutron stars. This phase transition alters the neutron star equation of state. Here, applying the scalar-tensor theories of gravity, we investigate the structure of neutron stars which are in the ferromagnetic phase. Considering the equation of state of ferromagnetic neutron matter with Skyrme-type interactions at zero temperature and using the scalar-tensor theories of gravity with sufficiently negative coupling constant, we explore the spontaneous scalarization in ferromagnetic neutron stars. In this regard, the mass versus the central density, the profiles of scalar field and mass and density, the central scalar field for different central densities, and the mass-radius relation of ferromagnetic neutron stars are presented. Moreover, we investigate the influences of the coupling constant on the scalarization of ferromagnetic neutron stars. The effects of the coupling constant on the critical densities of scalarization and the scalar charge of ferromagnetic neutron stars are also calculated. In addition, we study the maximum value of the coupling constant at which the
spontaneous scalarization takes place presenting the influence of the neutron matter equation of state on this critical value.
\end{abstract}
\pacs{21.65.-f, 26.60.-c, 64.70.-p}

\maketitle

\section{INTRODUCTION}

Ferromagnetic spin state in neutron stars is one of the most interesting subjects for both
nuclear and particle astrophysicists.
Different investigations verify that this ordering takes place in neutron stars \cite{Brownell,Rice,
Niembro,Marcos,Kutschera,Akhiezer,Maruyama,Beraudo,Eto,Hashimoto,Bernardos,Isayev4,Isayevj,Rios}.
The conditions for a ferromagnetic transition in neutron
stars in the framework of exact general-relativistic
equilibrium hydrodynamics and applying the relativistic equation of state (EOS) for interacting
hard-core nucleons have been presented \cite{Brownell}.
They showed that a ferromagnetic transition
in these stars is not inconsistent with stellar equilibrium in possible
superdense stars.
The ferromagnetic transition has been explored in a dense neutron matter system
employing the Landau's Fermi liquid function for a hard-sphere gas of neutrons \cite{Rice}.
Applying the relativistic Hartree-Fock approach to a pure neutron system with the densities
appropriate for neutron stars shows that the ferromagnetic transition is
specified by the Fock terms \cite{Niembro}.
Within a relativistic Hartree-Fock approach and using different combinations of mesons and couplings, it has been clarified that
a $\pi$-meson pseudoscalar coupling prefers the ferromagnetic phase in the pure neutron system \cite{Marcos}.
Studying the neutron matter with Skyrme forces verifies that with the parameters which provide a
very good parameterization of realistic neutron matter calculations, the ferromagnetic spin ordering
occurs in the neutron matter \cite{Kutschera}.
The Landau Fermi liquid theory confirms that the spontaneous magnetization takes place in dense degenerate neutron system
and this phase transition determines the neutron star density and magnetic field \cite{Akhiezer}.
Studying the highly dense nuclear matter with the relativistic
mean-field approach confirms that when the axial-vector interaction is negative enough, the
system holds ferromagnetism \cite{Maruyama}.
Employing the Hartree-Fock theory for a system of nucleons interacting through a central spin-isospin schematic
force and using the technique of the anomalous propagator, the critical
values of the interaction that leads to a transition to a ferromagnetic phase have been obtained \cite{Beraudo}.
The axial anomaly acting on the parallel layers of neutral
pion domain walls which spontaneously formed at high density in chiral nonlinear sigma model results in a ferromagnetic
phase at high density for degenerate neutron matter \cite{Eto}.
Noting the chiral effective model incorporating magnetic fields and the chiral anomaly, it has been shown that
with an axial vector meson condensation, there will be a possible realization of a QCD ferromagnetic phase and ferromagnetic
magnetars \cite{Hashimoto}.
The calculation in the
framework of a relativistic $\sigma+\omega+\pi+\rho$ Hartree-Fock approach verifies that the interaction of protons with
neutrons leads to the lower densities of spontaneous magnetization compared to the pure neutron matter \cite{Bernardos}.
Considering the presence of the protons with neutrons in a Hartree-Fock methed with Skyrme forces also leads to the
reduction of the density of ferromagnetism \cite{Isayev4}.
For asymmetric nuclear matter in the framework of a Fermi liquid theory with the Skyrme effective interaction,
the system experiences a phase transition to the spin
polarized state with the oppositely directed spins of neutrons
and protons, i.e. antiferromagnetic spin state \cite{Isayev4}.
However, considering the phase transition of symmetric nuclear matter in Fermi liquid theory with the Skyrme effective interaction confirms that the ferromagnetic spin state is preferable over the antiferromagnetic one \cite{Isayevj}.
The critical density of ferromagnetic transition in neutron matter with Skyrme-type interactions
decreases with temperature \cite{Rios}.

Scalar-tensor theories of gravity (STTs) which are among the extensions of general relativity (GR)
have been studied and developed by Fierz \cite{fierz}, Jordan \cite{jordan}, Brans and
Dicke \cite{brans}, Bergman \cite{Bergmann}, Nordtvedt \cite{Nordtvedt}, Wagoner \cite{Wagoner}, Damour
and Esposito-Farese \cite{damour}.
These theories which include one or more scalar fields coupled to matter, predict scalar gravitational wave that can
be detected by the laser interferometer \cite{Shibata94,Harada97}.
These theories of gravity have been extensively applied to investigate
the relativistic compact objects
\cite{damourPRL,Harada,Novak,beta-4.35,Novak2,Doneva,Ramazanoglu,Yazadjiev,sotani,Motahar,Mendes8,Kobayashi,Sotani8,Staykov,Doneva18}.
It has been confirmed that a wide class of STTs passes the weak-field gravitational
tests indicating nonperturbative strong-field deviations away from GR in neutron stars \cite{damourPRL}.
The instability in spherically symmetric stars in STTs is induced by the scalar field for some ranges of the
value of the first derivative of the coupling function \cite{Harada}.
In the gravitational collapse of neutron star toward a black hole, the amplitude of the
gravitational wave increases with  the value of the parameter of the coupling function \cite{Novak}.
Using a turning point method, the secular stability of neutron stars in STTs against spherically
symmetric perturbations has been studied \cite{beta-4.35}.
Considering spherical neutron star models in STTs, it has been shown that
with some conditions on the second derivative of the coupling function and on star's
mass, there exist two strong-scalar-field solutions as well as the usual weak-field
one \cite{Novak2}.
Solving the field equations describing the equilibrium of rapidly rotating neutron
stars in scalar-tensor theories of gravity shows that the deviations of the rapidly rotating
scalar-tensor neutron stars from the GR
solutions is significantly larger than in the static case \cite{Doneva}.
Studying the effect of a mass term in the spontaneous scalarization of neutron stars confirms that
this addition is a natural extension to the model that avoids the observational bounds on the STTs \cite{Ramazanoglu}.
Investigation of slowly rotating neutron stars in scalar-tensor theories with a massive gravitational
scalar fiels verifies that that mass, radius and
moment of inertia for neutron stars in massive scalar-tensor theories can differ drastically
from the pure GR solutions \cite{Yazadjiev}.
For lower sound velocity in the core of neutron stars, the maximum mass limit
of the scalarized neutron stars is larger than that of in GR, while for the stiff EOSs with high sound velocity, the maximum mass limit in GR is larger than the one in STT \cite{sotani}.
Applying different realistic equations of state, including pure nuclear matter, nuclear matter with hyperons, hybrid nuclear and
quark matter, and pure quark matter to study the scalarization of static and slowly rotating neutron
stars shows that the magnitude of the scalarization is correlated with the value of the gravitational potential at
the center of the star \cite{Motahar}.
The coupling
to additional degrees of freedom in massless STTs leads to new families
of modes in the spectrum of pulsating neutron stars, with no counterpart in GR \cite{Mendes8}.
Relativistic stars in degenerate higher-order STTs which evade the constraint
on the speed of gravitational waves imposed by GW170817 have been investigated \cite{Kobayashi}. Their results indicate that for high density stars, the mass-radius relation changes from the one in GR.
For the coupling constant confined to values provided by the astronomical
observations, the maximum compactness of neutron stars in GR is higher than the one in STT \cite{Sotani8}.
Applying scalar-tensor theory with massive field with self-interaction
term in the potential to the neutron star models demonestrates that the self-interaction
term suppresses the scalar field and large deviations from pure GR is
observed \cite{Staykov}.
Using a class of scalar-tensor theories of gravity indistinguishable from GR in the
weak field regime but with significant deviations in strong fields, it has been shown that
the maximum mass of a differentially rotating neutron star increases significantly
for scalarized solutions and such stars can reach larger angular momenta \cite{Doneva18}.
Regarding the above discussions, the EOS of ferromagnetic neutron matter can have significant effects on the
scalarized neutron stars. The main goal of this work is investigating the effects of the ferromagnetic neutron matter on the structure of neutron stars in STTs. Considering the ferromagnetic neutron matter with Skyrme-type interactions within the STT, we investigate the structure as well as the spontaneous scalarization of ferromagnetic neutron stars.

\section{  Formalism  }

We note a homogeneous system of $N$ neutrons with the spin-up number density $n_\uparrow$, spin-down number density $n_\downarrow$, and total number density $n=n_\uparrow+n_\downarrow$. The Skyrme-like effective interactions have the standard form \cite{chabanat,Rios},
\begin{eqnarray}
 V({{\verb"r"_1}},{\verb"r"_2})&=&t_0(1+x_0P^{\sigma})\delta({\verb"r"})\nonumber\\
 &+&\frac{1}{2}t_1(1+x_1P^{\sigma})({{\verb"K"{'}^2}}\delta({\verb"r"})+{{\delta({\verb"r"})\verb"K"^2}})
  +t_2(1+x_2P^{\sigma})\verb"K"{'}.\delta({\verb"r"})\verb"K"\nonumber\\ &+&\frac{1}{6}t_3(1+x_3P^{\sigma})
 [\rho(\verb"R")]^{\gamma}\delta({\verb"r"})+iW_0(\sigma_1+\sigma_2).[\verb"K"{'}\times\delta({\verb"r"})\verb"K"],
\end{eqnarray}
with $\verb"r"=\verb"r"_1-\verb"r"_2$, $R=(\verb"r"_1+\verb"r"_2)/2$, $\verb"K"=(\nabla_1-\nabla_2)/2i$, and $P^{\sigma}=(1+\sigma_1.\sigma_2)/2$. Besides, $\verb"K"$, the relative momentum acts on
the right and $\verb"K"{'}$ is its conjugate and acts on the left. Moreover, $t_0$, $t_1$, $t_2$, $t_3$, $x_0$, $x_1$, $x_2$, $x_3$, $\gamma$, and $W_0$ are the Skyrme-force parameters. The total energy of neutron matter per particle is calculated by $E/N=\langle\psi|H|\psi\rangle/N$ through the Hartree-Fock method in which $\psi$ and $H$ denote the wave function and the Hamiltonian of system, respectively. For the non-ferromagnetic neutron matter (NFM), the energy per particle is as follows \cite{chabanat},
\begin{eqnarray}
 E_{NFM}/N&=&\frac{3}{5}\frac{\hbar^2}{2m}(3\pi^2 n)^{2/3}+\frac{1}{4}t_0(1-x_0)n+\frac{1}{24}t_3(1-x_3)n^{\gamma+1}+\frac{3}{40}(3\pi^2)^{2/3}\Theta n^{5/3},
\label{unpol}
\end{eqnarray}
with $\Theta=t_1(1-x_1)+3t_2(1+x_2)$. We also consider a system of polarized neutron matter with the spin-up number density $n_\uparrow$, spin-down number density $n_\downarrow$, and spin polarization parameter $\Delta=(n_\uparrow-n_\downarrow)/n$. The energy per particle for ferromagnetic (FM) neutron matter is given by \cite{Rios},
\begin{eqnarray}
 E_{FM}/N&=&\frac{\hbar^2}{2m}\frac{\tau_\uparrow+\tau_\downarrow}{n}+\frac{1}{4n}[2t_2(1+x_2)]
 [\tau_\uparrow n_\uparrow+\tau_\downarrow n_\downarrow]\nonumber\\ &+&\frac{1}{4n}[t_1(1-x_1)+t_2(1+x_2)]
 [\tau_\uparrow n_\downarrow+\tau_\downarrow n_\uparrow]\nonumber\\ &+&\frac{1}{n}[t_0(1-x_0)+\frac{1}{6}t_3(1-x_3)n^\gamma]n_\uparrow n_\downarrow,
 \label{pol}
\end{eqnarray}
with
\begin{eqnarray}
\tau_{\sigma}=\frac{3}{10}(3\pi^2 n)^{2/3}n(1 \pm \Delta)^{5/3},
\end{eqnarray}
in which $+$ and $-$ denote the up and down spin projections. It is possible to calculate
the equilibrium value of the spin polarization parameter, i.e. $\Delta_{min}$, at each number density $n$ from Eq. \eqref{pol}.
By the equilibrium value, we mean the value at which the energy is minimum.
Besides, the energy at each $\Delta_{min}$ can be calculated using Eq. \eqref{pol}. Here, we employ the SLy230a model to describe the neutron matter EOS \cite{chabanat}. At lower densities, $\Delta_{min}$ is equal to zero and neutron matter is fully unpolarized. At a density about $5.40 \rho_0$ in which $\rho_0=1.66\times10^{17}kg/m^3$, this parameter spontaneously increases and it reaches to 1 at a density about $10.41 \rho_0$.
Neutron matter with the densities between $5.40 \rho_0<\rho<10.41 \rho_0$ is partially polarized. Besides, with the densities higher than  $10.41 \rho_0$, the neutron matter becomes fully polarized and it is in ferromagnetic state.
Applying the first law of thermodynamics, i.e. $P=n^2(\frac{\partial (E/N)}{\partial n})$, the EOSs for the
NFM and FM neutron matter are calculated. Fig. \ref{1-eos} shows the EOSs of non-ferromagnetic and ferromagnetic neutron matter in Sly230a model.
\begin{figure}[h!]
		{\includegraphics[scale=0.90]{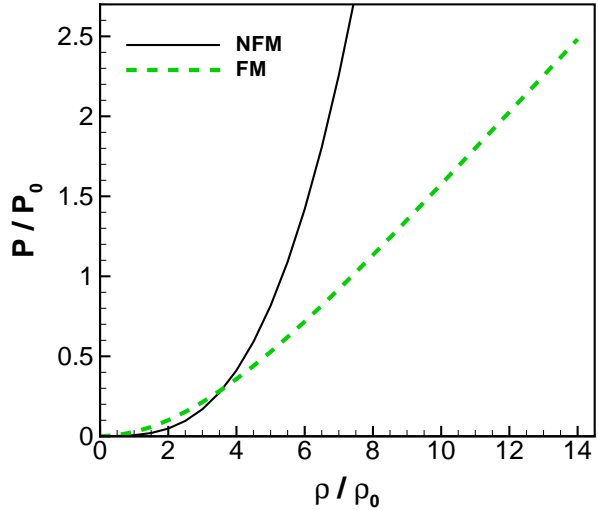}
		\label{1-eos}}
		\caption{Pressure, $P$, of the non-ferromagnetic (NFM) and ferromagnetic (FM) neutron
 matter in SLy230a model versus the density, $\rho$, with $P_0=5\times10^{34}kg/ms^2$ and $\rho_0=1.66\times10^{17}kg/m^3$.}
	\label{1-eos}
\end{figure}
For both NFM and FM neutron matter, the pressure grows by increasing the density. At most values of $\rho$, the rate at which the pressure increases is higher for NFM neutron matter. Therefore, the EOS of FM neutron matter is softer than NFM one. As wee see in the following, the softening of the EOS affects the properties of the scalarized neutron stars, i.e. mass, radius, and spontaneous scalarization. The main goal of this work is investigating the effects of neutron matter ferromagnetism on the properties of neutron stars in STTs.

To describe the neutron stars in STTs, we start with the spacetime line element in the Einstein frame for a spherical symmetry static neutron star as follows \cite{mendes},
\begin{equation}
ds^2 = - N(r)^2 dt^2 + A(r)^2 dr^2 + r^2
(d\theta^2 + \sin^2\theta d\phi^2),
\end{equation}
in which $N(r)$ and $A(r)$ are the metric functions. The function $A(r)$ is related to the mass by $A(r) := [1-2 m(r)/r ]^{-1/2}$. We also consider the neutron star as a perfect fluid with stress-energy tensor as follows,
\begin{align}
	\tilde{T}^{\mu\nu} = \tilde{\epsilon} \tilde{u}^\mu \tilde{u}^\nu + \tilde{p} (\tilde{g}^{\mu\nu} + \tilde{u}^\mu \tilde{u}^\nu),
\end{align}
with the fluid's total energy density $\tilde{ \epsilon}$, pressure $\tilde{p}$, and 4-velocity $\tilde{u}$. It should be noted that tilde presents the quantities in Jordan frame. Besides, we denote the physical metric or Jordan metric by $\tilde{g}_{\mu\nu}:=a(\phi)^2g_{\mu\nu}$ in which $a(\phi)$ and $\phi$ are the coupling function and scalar field, respectively. For the coupling function $a(\phi)$, we consider two models as follows \cite{damourPRL,mendes},
\begin{align}
\textrm{{\bf Model 1} (M1)}:\quad &a(\phi) =[ \cosh \left(\sqrt{3} \beta (\phi-\phi_0)\right)]^\frac{1}{3\beta}, \nonumber \\
& \alpha(\phi) = \frac{1}{\sqrt{3}} \tanh (\sqrt{3} \beta (\phi-\phi_0)), \label{eq:alpha1} \\
\textrm{{\bf Model 2} (M2):}\quad &a(\phi) = e^{\frac{1}{2}\beta (\phi-\phi_0) ^2}, \nonumber \\
& \alpha(\phi) = \beta(\phi-\phi_0), \label{eq:alpha2}
\end{align}
in which $\beta$ is the coupling constant and we assume $\phi_0=0$. In addition, $\alpha(\phi) := \frac{d\ln a(\phi)}{d\phi}$.
It has been verified that a nonperturbative
amplification mechanism of the coupling
strength of the scalar field exists when the logarithm of the
coupling function has a sufficiently negative curvature
around $\phi_0$ \cite{damourPRL}. They showed the existence of strong-field deviations from GR in STTs with $\beta \lesssim -4$.
Besides, the negative values of $\beta$ result in the
scalar field nonlinearities and these reinforce the naturally
attractive character of scalar interactions \cite{damour1996}.
We know that the binary pulsar experiments set constraints on the value of the coupling constant i.e. $\beta > -4.8$ \cite{Freire}.
Moreover, the pulsar-white dwarf binary PSR J0348+0432, put a bound on the coupling constant $\beta\geq-4.5$ \cite{Freire}.
These constraints are in the massless scalar field case. However, adding a mass term
to the scalar field potential results in the extension to the model that avoids these observational bounds  \cite{Ramazanoglu}. In fact, the coupling constant can be much smaller than $-4.5$ for massive STTs \cite{Yazadjiev}.
We apply different values for the coupling constant which are lower and larger than
the lower limit set by the binary pulsar experiments. Consideration of the values smaller than the lower limit
from the observations is justified because in the present study which is the first investigation of the scalarized
FM neutron stars, we are interested on the effects of the coupling constant on the scalarization of these stars.
In addition, the calculations with $\beta<-4.5$ in the massless case give the upper limit for the deviations from GR in the massive case \cite{Doneva18}.
We also calculate the maximum value of the coupling constant at which the spontaneous scalarization takes place in
NFM  and FM neutron stars. This determines the influence of the neutron matter EOS on this critical value. This value has previously calculated using a polytropic equation of state to be $-4.35$ in the nonrotating star \cite{beta-4.35} and $-3.9$ in rapidly rotating stars \cite{Doneva}.

By applying some calculations, the field equations in STTs lead to the following equations which describe the structure of neutron stars in STTs \cite{mendes},
\begin{align}
&\frac{d m}{dr} = 4\pi r^2 a^4 \tilde{\epsilon} + \frac{r}{2} (r-2m) \Big(\frac{d\phi}{dr}\Big)^2 \label{eq:dm},\\
&\frac{d \ln N}{dr} = \frac{4\pi r^2 a^4 \tilde{p}}{r - 2m} +\frac{r}{2} \Big(\frac{d\phi}{dr}\Big)^2 + \frac{m}{r(r-2m)} \label{eq:dn}, \\
&\frac{d^2\phi}{dr^2} = \frac{4\pi r a^4}{r-2m} \! \left[ \alpha (\tilde{\epsilon} - 3\tilde{p}) + r (\tilde{\epsilon} - \tilde{p}) \frac{d\phi}{dr} \right ]\! -\frac{2(r-m)}{r(r-2m)} \frac{d\phi}{dr} \label{eq:dphi}, \\
&\frac{d\tilde{p}}{dr} = -(\tilde{\epsilon} + \tilde{p}) \left[  \frac{4\pi r^2 a^4 \tilde{p}}{r-2m} \! + \! \frac{r}{2} \Big(\frac{d\phi}{dr}\Big)^2 \!\! + \! \frac{m}{r(r-2m)} \! + \! \alpha \frac{d\phi}{dr} \right], \label{eq:dp}\\
&\frac{dm_b}{dr}=\frac{4\pi r^2a^3\tilde{\rho}}{\sqrt{1-\frac{2m}{r}}}\label{eq:dm_b},
\end{align}
in which $m_b$ denotes the baryonic mass. These are the generalized Tolman-Oppenheimer-Volkoff equations in STTs.
The boundary conditions to solve these equations together with the neutron matter EOS are as follows,
\begin{align}
&m(0) =m_b(0)= 0, \quad \lim_{r\to\infty}N(r) = 1,\quad \phi(0)=\phi_c, \quad \lim_{r\to\infty}\phi(r) = 0, \nonumber \\
&\frac{d\phi}{dr}(0) = 0, \qquad \tilde{p}(0) = p_c, \qquad \tilde{p}(R_s) = 0, \label{eq:bc}
\end{align}
where $R_s$ is the radius of the star. Using a fourth-order Runge-Kutta algorithm \cite{Mathematica}, we integrate the above equations.
The integration is done with the boundary conditions at $r=0$. In addition, with a guess for the scalar field at the center, i.e. $\phi(0)=\phi_c$, we do the iteration on $\phi_c$ such that the following condition satisfies \cite{damourPRL,mendes},
\begin{equation} \label{eq:constraint on central of scalar field}
\phi_s  + \frac{2 \psi_s}{\sqrt{\dot{\nu}_s^2+4\psi_s^2}} \textrm{arctanh} \left[ \frac{\sqrt{\dot{\nu}_s^2 +4\psi_s^2}}{\dot{\nu}_s +2/R_s} \right] = 0.
\end{equation}
In the above equation, the subscript s denotes the quantities on the surface of star.
Besides, $\psi_s := (d\phi/dr)_s$ and $\dot{\nu}_s := 2(d\ln N/dr)|_s = R_s \psi_s^2 + 2 m_s/[R_s(R_s-2m_s)]$. Moreover, the ADM mass, $M_{ADM}$, and scalar charge, $\omega$, are given by \cite{damourPRL,mendes},
\begin{align}
M_{ADM} &= \frac{R_s^2 \dot{\nu}_s}{2} \left( 1-\frac{2m_s}{R_s} \right)^\frac{1}{2}
 \exp \left[ \frac{-\dot{\nu}_s}{\sqrt{\dot{\nu}_s^2+4\psi_s^2}} \textrm{arctanh} \left( \frac{\sqrt{\dot{\nu}_s^2+4\psi_s^2}}{\dot{\nu}_s +2/R_s} \right) \right], \\
\omega & = - 2 M_{ADM} \psi_s/\dot{\nu}_s.
\end{align}
It should be mentioned that the scalar charge is introduced through the asymptotic
behavior of the scalar field with $r\rightarrow\infty$ as follows \cite{damour},
\begin{align}
\phi(r)=\omega/r+\mathcal{O}(1/r^2).
\end{align}

\section{RESULTS and DISCUSSION}\label{RESULTS}
\subsection{Mass versus the Central Density}\label{subsec:M-rho}
 Fig. \ref{2-m-rho} demonstrates the NFM and FM neutron star mass as a function of the central density for the STT and GR in two models (M1 and M2) of coupling function for different values of the coupling constant.
\begin{figure}[h!]
	{\includegraphics[scale=0.67]{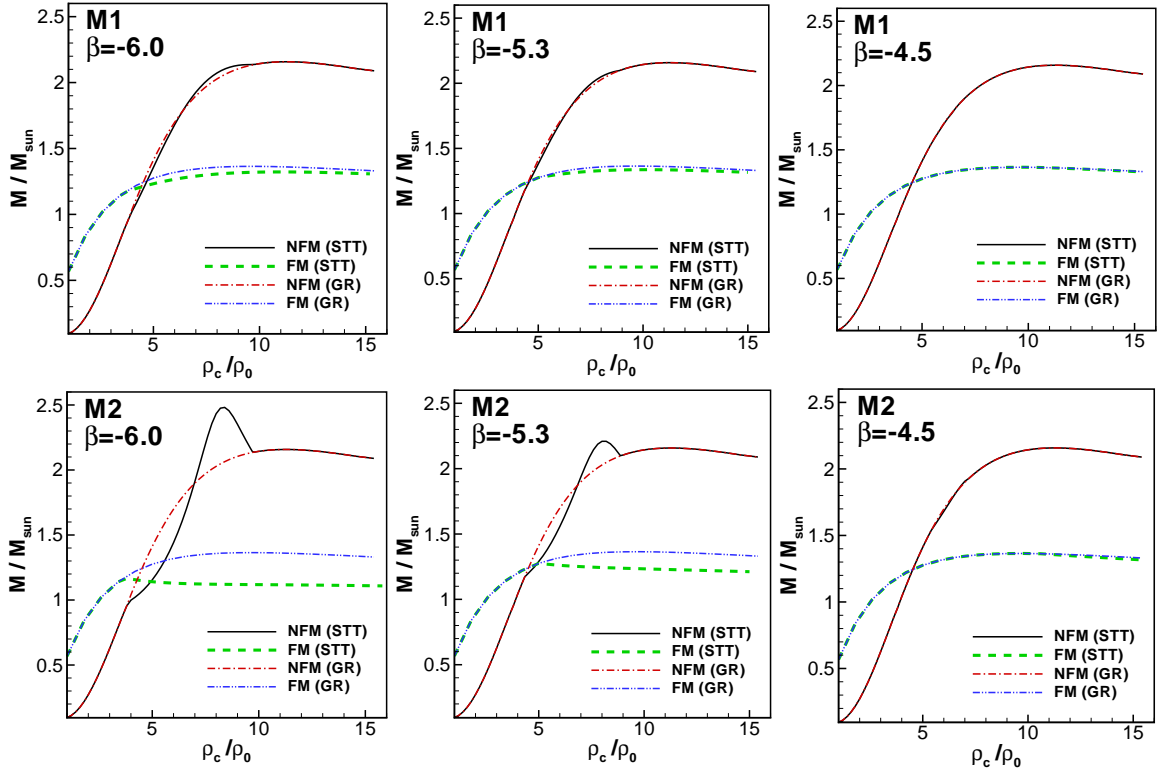}
		\label{2-m-rho}	}
		\caption{Neutron star mass, $M,$ versus the central density, $\rho_c,$ for
		 NFM and FM neutron stars in STT and GR applying M1 and M2 with different values of the coupling constant, $\beta$.  }
	\label{2-m-rho}
\end{figure}
In M1 for different values of the coupling constant and M2 for higher values of the coupling constant, the neutron star mass increases by increasing the density to a special value of the central density. For the densities higher than that special value, the neutron star mass is constant. This special value of the central density is greater for NFM neutron stars. At lower densities, the mass of FM neutron stars is greater than that of NFM ones. But at higher densities, the mass of NFM neutron stars is higher than that of FM stars. This is due
to this fact that the EOS of NFM neutron matter is stiffer than the EOS of FM neutron matter. In addition, in M1 for different values of $\beta$ and in M2 for higher values of $\beta$, for both NFM and FM neutron stars, the results of GR and STT are nearly equal. This is while the results of STT and GR are different for lower values of $\beta$. For the cases that the results of GR and STT are different, the neutron stars are scalarized. In M2 with lower values of $\beta$, the mass of the stars with lower central densities in STT is smaller than GR.
However, for some stars with higher densities, the mass in STT is greater than GR. As we see in the following, this fact that the result of STT how is different from GR is related to the variation of the central scalar field respect to the central density.
For NFM neutron stars with high densities, the results of STT and GR are the same. Therefore, there is no scalarized NFM neutron stars with high central density. However, for the FM neutron stars with densities greater than a special value, the results of GR and STT are not the same. This difference continues to highest values of the density considered in this work, i.e. $\rho_c=14\rho_0$. Therefore, even the FM neutron stars with high densities are scalarized unlike NFM stars. In fact, this phenomenon (scalarization of FM neutron stars with high densities) is the main distinction of NFM and FM neutron stars.
For the lower values of $\beta$, the deviation of STT from GR is more considerable. Besides, the difference of the results in STT and GR is more significant in M2 compared to M1.
\begin{table}[h!]
	\begin{center}
		\begin{tabular}{cccc}
			\hline
			\hline
 & & Maximum Mass ($M_{\bigodot}$)& \\	
\hline
Model & $\beta$& STT (NFM) & STT (FM)\\	
\hline
{1}&$-6.0$ &2.15 &1.32 \\
&$-5.3$ &2.09 &1.33 \\
&$-4.5$ &1.93 &1.36 \\
\hline\hline
{2}& $-6.0$&2.48 &1.16 \\
&$-5.3$ &2.20 &1.27 \\
&$-4.5$ &1.93 &1.36 \\
\hline\hline			
			
		\end{tabular}
		
	\end{center}
	\caption{Maximum mass of NFM and FM neutron stars in STT for different values of the coupling constant in two models. Besides, the maximum mass for NFM and FM neutron stars in GR is $2.15\ M_{\bigodot}$ and $1.36\ M_{\bigodot}$, respectively.
	}
\label{table1:maximum mass}
\end{table}
\begin{table}[h!]
	\begin{center}
		\begin{tabular}{cccc}
			\hline
			\hline
			 & & Maximum Compactness& \\	
			 \hline
			Model & $\beta$& STT (NFM) & STT (FM)\\	
			\hline
			{1}&$-6.0$ &0.30 &0.16 \\
			&$-5.3$ &0.28 &0.16 \\
			&$-4.5$ &0.25 &0.17 \\
			\hline\hline
			{2}& $-6.0$&0.29 &0.13 \\
			&$-5.3$ &0.28 &0.15 \\
			&$-4.5$ &0.25 &0.16 \\
			\hline\hline			
					\end{tabular}
		
	\end{center}
	\caption{Maximum compactness of NFM and FM neutron stars in STT for different values of the coupling constant in two models. Besides, the maximum compactness for NFM and FM neutron stars in GR is $0.33$ and $0.17$, respectively.}
	\label{table2:maximum comp}
\end{table}
In Table \ref{table1:maximum mass}, we present the maximum mass for NFM and FM stars. According to these results, for NFM neutron stars in two models, the maximum mass decreases by increasing the coupling constant. The difference of maximum mass in two models decreases by increasing $\beta$. In FM stars unlike the NFM ones, the maximum mass in two models grows with the increase of $\beta$. So we can conclude that the effects of $\beta$ on the maximum mass depend on the EOS of neutron matter. With stiffer EOS (NFM neutron matter), the maximum mass reduces as $\beta$ grows. However, with softer EOS (FM neutron matter), the maximum mass increases by increasing $\beta$. In addition, for NFM neutron stars, the maximum mass in M2 is greater than M1.
However, this difference of two models is opposite for the FM stars. It should we noted that the maximum mass of FM stars in STT is always smaller than or equal to one in GR. In both gravities, the maximum mass of FM stars is lower than the NFM ones. This is due to the fact that the EOS of FM neutron matter is softer than the NFM one. Table \ref{table2:maximum comp} also gives the maximum compactness of NFM and FM neutron stars.
By increasing $\beta$, the maximum compactness of stars in STT, reduces for NFM stars while it grows for FM ones. Moreover, for almost all NFM and FM neutron stars, the maximum compactness in STT
is lower than the one in GR. This effect is in agreement with the one reported in Ref. \cite{Sotani8}.

\subsection{Profiles of Scalar Filed, Mass, and Density}
Fig. \ref{3-phi-r} shows the profile of scalar field for NFM and FM neutron stars in STT with two models. The value of scalar field is nonzero in each point of the stars. Moreover, for different values of $\beta$, at each distance to the center of star, M2 predicts higher values for the scalar field compared to M1.
\begin{figure}[h]
{\includegraphics[scale=0.67]{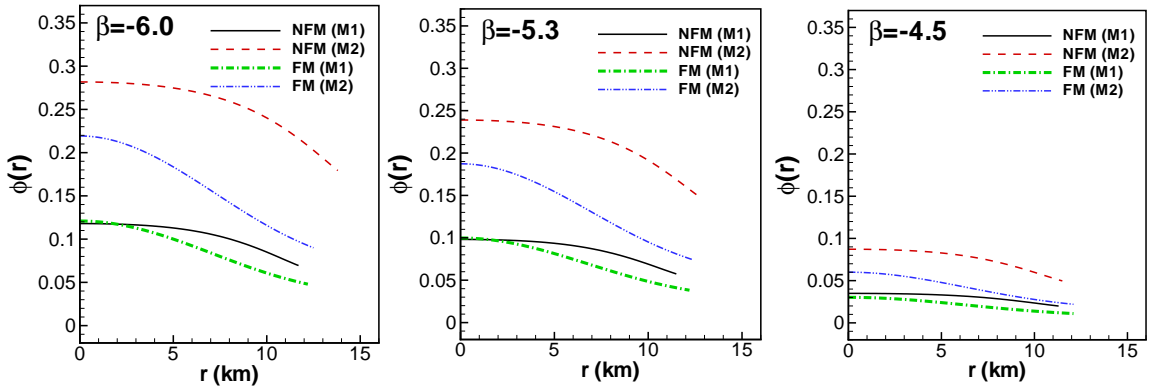}
\label{3-phi-r}	}
		\caption{Profile of scalar field in NFM and FM neutron stars for two models with different values of the coupling constant, $\beta$. The values of the central density are $\rho_c=7\rho_0$ and $\rho_c=12\rho_0$ for
the NFM and FM neutron stars, respectively.}
	\label{3-phi-r}
\end{figure}
This is because in M2, the coupling of scalar field to metric is more significant compared to M1. The rate at which the scalar field reduces is higher in M2.
Our calculations verify that by increasing $\beta$, the scalar field in neutron star decreases. This is due to the fact that for lower values of $\beta$, the coupling of scalar field to metric is more significant. In addition, the scalar field for FM neutron star is smaller than the NFM one. Therefore, with the softer EOS, the magnitude of the scalar field is smaller than the one with the stiffer EOS. The difference between the profile of scalar field in two models is more considerable for NFM neutron stars. Moreover, with different values of $\beta$, for both NFM and FM neutron stars, the difference of M1 and M2 is more important in the center of star compare to its surface.

Fig. \ref{4-m-r} presents the profile of mass for NFM and FM neutron stars in STT with two models.
\begin{figure}[h]
	{\includegraphics[scale=0.67]{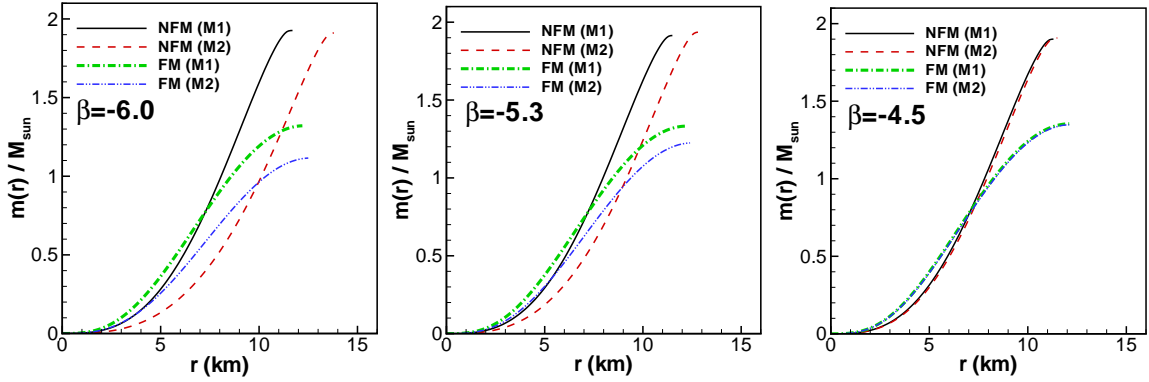}
		\label{4-m-r}	}
		\caption{Same as Fig. \ref{3-phi-r} but for the profile of mass.}
\label{4-m-r}
\end{figure}
According to Fig. \ref{4-m-r}, the mass profiles in two models are different for both NFM and FM neutron stars. This difference is more significant for NFM stars. With higher values of $\beta$, the profiles approach to each other. The mass profile of FM stars in two models is lower than the NFM one. This is due to the fact that the FM EOS is softer than NFM one.
Fig. \ref{5-rho-r} shows the profile of density for NFM and FM neutron stars in STT with two models. For the profile of density, the difference between two models is more considerable at lower coupling constants.
This difference is nearly negligible for FM stars.
 \begin{figure}[h]
 	{\includegraphics[scale=0.67]{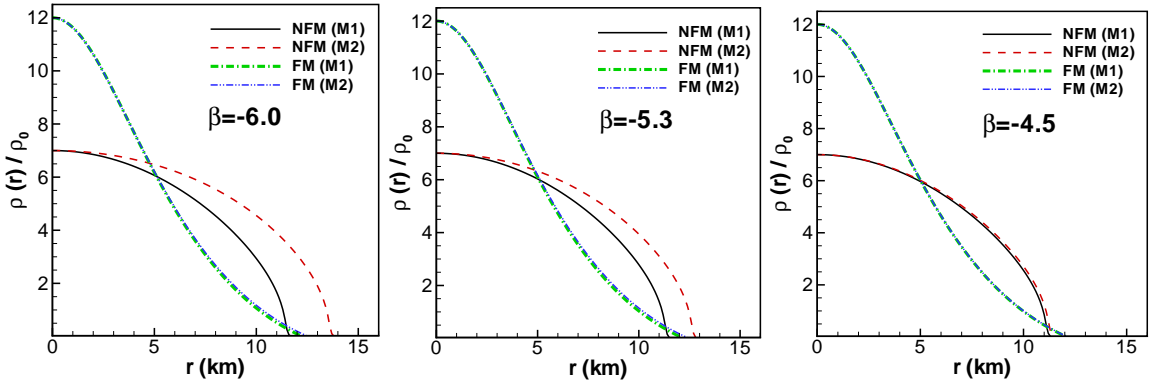}
 		\label{5-rho-r}	}
 	 	\caption{Same as Fig. \ref{3-phi-r} but for the profile of density.}
 	\label{5-rho-r}
 \end{figure}
Fig. \ref{5-rho-r} confirms that by increasing $\beta$, the density becomes equal to zero at smaller distances to the center of star (i.e. stars with smaller radii). Moreover, at each coupling constant, M2 predicts larger radii for NFM stars. It should be noted that for FM neutron stars, the effect of coupling constant on the radius is not considerable.

 \subsection{Central Scalar Field versus the Central Density}
\begin{figure}[h]
\subfigure{\includegraphics[scale=0.67]{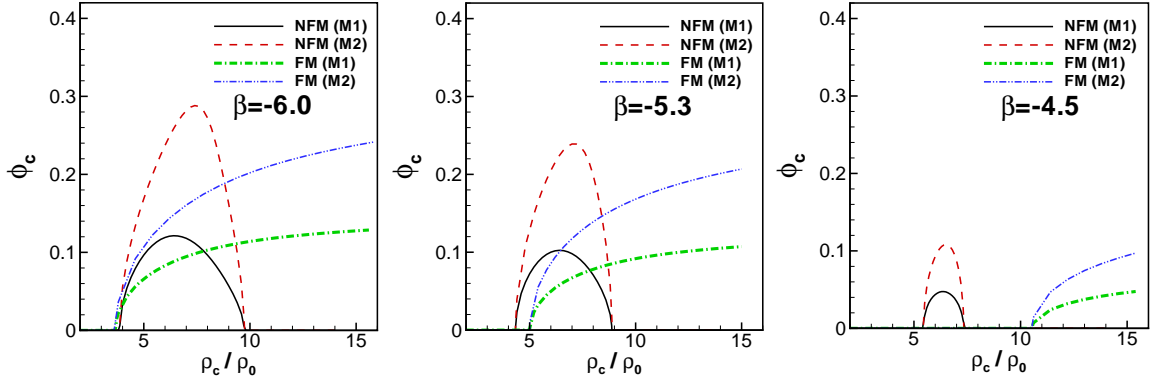}
}	\caption{Central scalar field, $\phi_c$, versus the central density, $\rho_c$, for NFM and FM neutron stars applying M1 and M2 with different values of the coupling constant, $\beta$. Note that in GR we have $\phi_c=0$.}
	\label{6-phi-rho}
\end{figure}

Fig. \ref{6-phi-rho} shows the central scalar field versus the central density for NFM and FM neutron stars. For both NFM and FM stars considering two models with all values of the coupling constant, the central scalar field is zero at lower densities. For stars with zero central scalar field, the solutions of GR and STT are equal. Considering both NFM and FM stars, the scalar field increases by increasing the density from a certain value and therefore the spontaneous scalarization takes place. However, the densities at which the scalarization takes place are different for NFM and FM neutron stars. In this work, we denote the first critical density of scalarization by $\rho_{cr1}$.
The GR and STT solutions are different at nonzero scalar fields. In both models for all values of $\beta$, the scalar field of
NFM stars becomes zero at a value of density (second critical density of scalarization, $\rho_{cr2}$). Moreover, for these stars, the scalar field remains zero up to high densities. Consequently, the high density NFM neutron stars are not scalarized. However, for FM neutron stars,
the scalar field increases monotonically by increasing the density and it does not become zero even at high densities.
Therefore, the high density FM neutron stars are also scalarized. This is the main difference of NFM and FM stars. This phenomenon is related to the one explained in Fig. \ref{2-m-rho}. In fact, the difference of the GR and STT solutions up to high densities for the mass verses the density in FM stars is a result of the nonzero scalar field in these stars. Therefore, the EOS of star affects its scalarization.
It can be seen from Fig. \ref{6-phi-rho} that with lower values of $\beta$, the first critical densities of scalarization for NFM and FM stars are more close to each other.
Besides, for lower values of $\beta$, the ranges of density at which the NFM and FM stars
are scalarized have more overlap with each other. For NFM stars in two models, the value of $\rho_{cr1}$ increases as the coupling constant grows, in agreement with the result of Ref. \cite{Sotani8}. In addition, for these stars, $\rho_{cr2}$ decreases by increasing $\beta$. It can be concluded that the range at which the NFM stars are scalarized
is larger for lower values of $\beta$. With lower values of the coupling constant, $\rho_{cr1}$ in FM stars is smaller than NFM ones. Therefore, with lower values of $\beta$, lower density FM stars can be also scalarized.
We can found that the range of scalarization in FM neutron stars is greater than NFM one.
In fact, softening of EOS leads to a larger range of scalarization. This effect agrees with the result reported in Ref. \cite{Salgado}. Another result of Fig. \ref{6-phi-rho} is that by increasing the coupling constant, the maximum value of the central scalar field decreases. For that reason, neutron stars are more scalarized with lower values of $\beta$. This result has also reported in Refs. \cite{damourPRL,Novak,Sotani8}. Fig. \ref{6-phi-rho} also verifies that M2 predicts higher values of the scalar field compared to M1. Therefore, the stars are more scalarized in M2.

\subsection{Mass-Radius Relation}
\begin{figure}[h]
	{\includegraphics[scale=0.67]{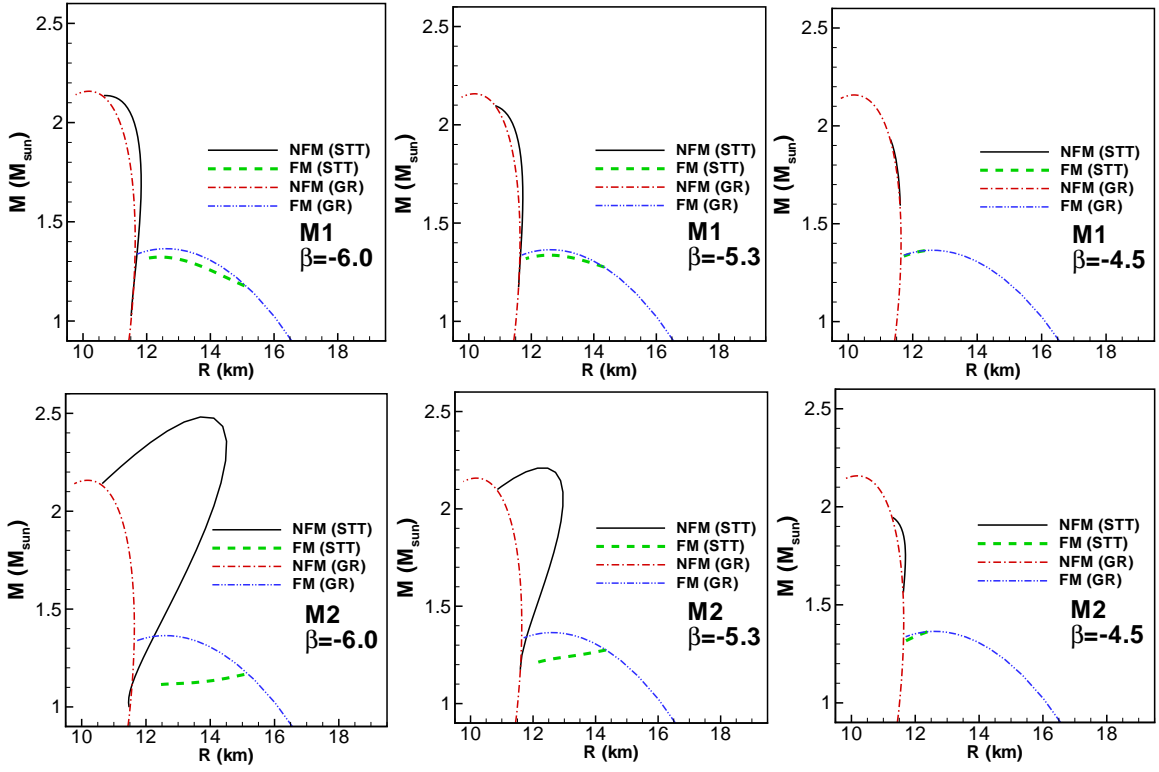}
	}	\caption{Mass versus the radius for
		 NFM and FM neutron stars in STT and GR applying M1 and M2 with different values of the coupling constant, $\beta$. The lines which show the allowed region for neutron stars are also presented.}
	\label{7-M-R}
\end{figure}

Fig. \ref{7-M-R} presents the mass-radius relation for NFM and FM neutron stars in STT and GR considering two models with different values of $\beta$.
The mass-radius relation of neutron stars which are not scalarized is equal in STT and GR.
With different values of $\beta$ in M1 and with high values of $\beta$ in M2, for NFM and FM neutron stars, the smaller stars have larger masses. Besides, for that conditions, the bigger stars have lower masses. In fact, they are gravitationally bound stars. Fig. \ref{7-M-R} shows that in two models for neutron stars, the deviation of STT from GR is more significant with lower values of $\beta$. According to Fig. \ref{7-M-R}, for massive NFM neutron stars, the results of STT and GR are equal. However, for massive FM neutron stars, these results are not the same. This is due to the fact that high density FM stars unlike the NFM ones are scalarized (see Fig. \ref{6-phi-rho}). In addition, this deviation in M2 is more considerable than M1. Because M2 predicts more scalarization for neutron stars compared to M1. For lower values of $\beta$, the deviation of STT from GR in the mass-radius relation takes place in a greater region. This is a result of the fact that with lower values of $\beta$, the range of scalarization is bigger (see Fig. \ref{6-phi-rho}). For scalarized NFM stars (specially with $\beta=-6.0$ in M2), the massive stars are bigger while the lower mass ones are smaller.
This means that the slop of mass-radius relation for these stars is different from the one in GR. In fact, in these cases, the stars are self bound. However, the scalarized FM neutron stars are still gravitationally bound. It is clear from Fig. \ref{7-M-R} that the deviation of STT from GR is more significant for NFM stars compared to FM ones. Therefore, it can be concluded that the EOS of neutron matter affects the amount of deviation of
STT from GR. The scalarized NFM neutron stars are larger in size compared to the GR solutions. In addition, for FM neutron stars, the mass of scalarized stars is lower than the stars in GR. This is similar to the result of Ref. \cite{mendes}.

\subsection{Critical Density of Scalarization}
In this part, we are going to investigate the effects of the coupling constant on the critical density of scalarization for NFM and FM stars. Fig. \ref{9-crde} gives the first and second critical densities for different stars in M1 versus the coupling constant. For both NFM and FM stars, the first critical density, $\rho_{cr1}$, increases as the coupling constant grows. This means that for higher values of $\beta$, the stars become scalarized
at higher densities. It should be noted that our results confirm that the critical density of scalarization does not depend on the coupling function model. It is clear from Fig. \ref{9-crde} that for the most values of $\beta$,
the first critical density in FM stars is higher than the one in NFM stars. Moreover, the rate at which $\rho_{cr1}$ grows with $\beta$ is greater for FM stars. The effects of the spin polarization of neutron matter on $\rho_{cr1}$ is more significant when the coupling constant is higher.
\begin{figure}
	\begin{center}
	{\includegraphics[scale=0.9]{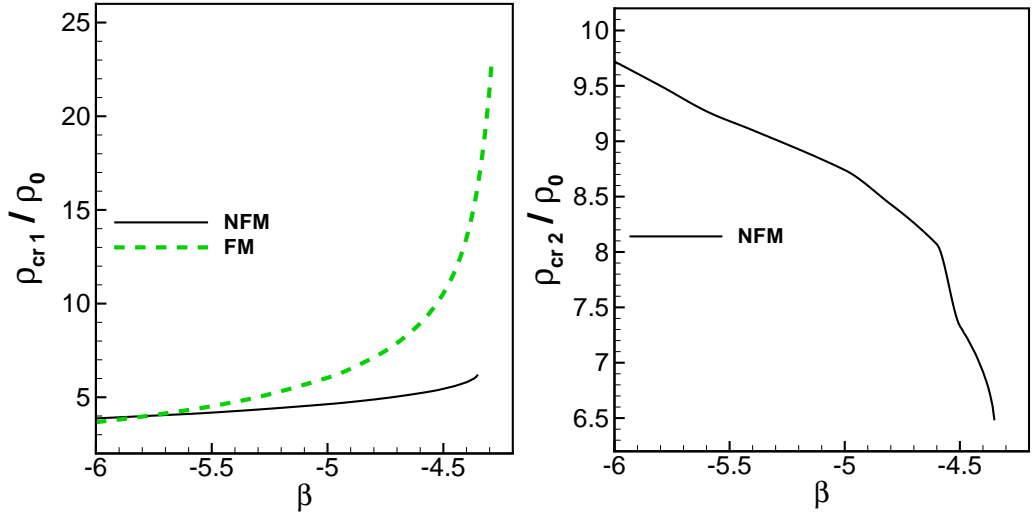}
\label{9-crde}}
\caption{First, $\rho_{cr1}$, and second, $\rho_{cr2}$, critical density of scalarization in NFM and FM neutron stars in M1 versus the coupling constant, $\beta$.}
	\label{9-crde}
		\end{center}
\end{figure}
Fig. \ref{9-crde} also shows the second critical density of scalarization for NFM stars versus $\beta$.
It should be mentioned that since the FM neutron stars are scalarized up to high densities, the second critical density
of scalarization is not defined for the FM stars. In NFM stars, the second critical density, $\rho_{cr2}$, decreases as $\beta$ grows.
Regarding the increase of $\rho_{cr1}$ with $\beta$, it is possible to conclude that for NFM neutron stars, the range of density at which the scalarization takes place decreases as the coupling constant grows. Therefore, the solution of STT approaches to GR when $\beta$ increases.

To explore how the neutron matter EOS affects the critical value of $\beta$, i.e. the maximum value of the coupling constant at which the scalarization takes place, we have presented the critical densities of scalarization in a smaller range of $\beta$ in Fig. \ref{9-crde-2}. It is clear that in the case of FM neutron stars, the maximum value of the coupling constant is higher compared to the NFM neutron stars. Therefore, in FM neutron stars, the range of the coupling constant at which the stars are scalarized is more extended. Our results confirm that the critical values of $\beta$ are $-4.35$ and $-4.29$ for NFM and  FM stars, respectively. According to our results, for NFM neutron stars with $\beta=-6.0$ and $\beta=-4.35$, the first critical densities are $\rho_{cr1}=3.87\rho_0$ and $\rho_{cr1}=6.21\rho_0$, respectively. Besides, with $\beta=-6.0$ and $\beta=-4.35$, the second critical densities are $\rho_{cr2}=9.72\rho_0$ and $\rho_{cr2}=6.48\rho_0$, respectively.
Moreover, for FM stars with $\beta=-6.0$ and $\beta=-4.29$, the values of the first critical density are $\rho_{cr1}=3.67\rho_0$ and $\rho_{cr1}=23.09\rho_0$, respectively.
\begin{figure}
	\begin{center}
	{\includegraphics[scale=0.9]{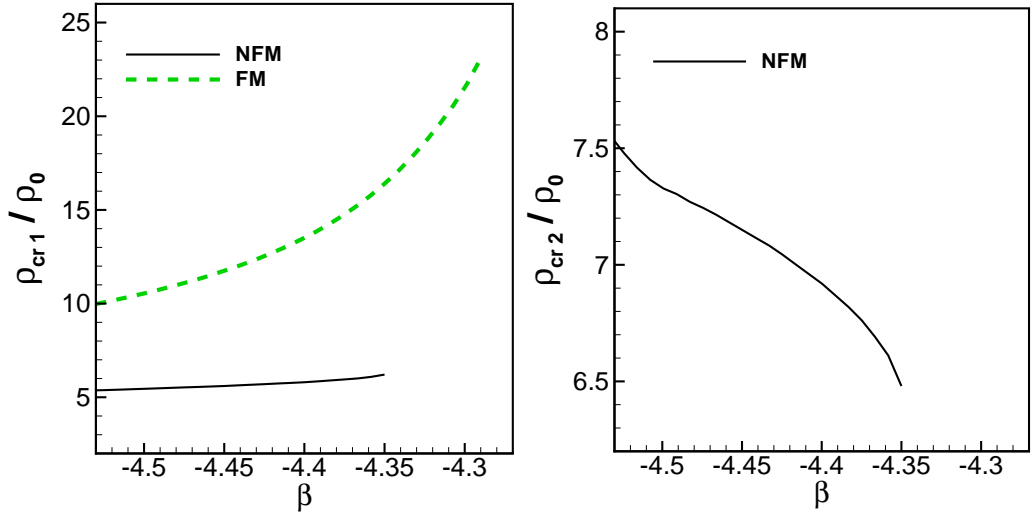}
\label{9-crde-2}}
\caption{Same as Fig. \ref{9-crde} but for a smaller range of the coupling constant.}
	\label{9-crde-2}
		\end{center}
\end{figure}

\subsection{Scalar Charge in Ferromagnetic and Non-Ferromagnetic Neutron Stars}
Fig. \ref{10-scalarcharge-M} presents the scalar charge versus the mass for NFM and FM neutron stars.
For both NFM and FM low mass stars with all values of $\beta$ in two models, the scalar charge is zero.
For all NFM neutron stars with the masses higher than a special value, the scalar charge grows when the mass increases.
For the NFM neutron stars, the mass at which the scalar charge becomes nonzero increases with the increase of $\beta$.
This is in agreement with the result of Ref. \cite{damour1996}.
For NFM neutron stars with high masses, the scalar charge again becomes zero. The range of mass with nonzero scalar charge is precisely the range at which the central scalar field is nonzero (see Fig. \ref{6-phi-rho}).
In fact, the more scalarized the neutron star, the more amount of scalar charge for star. Fig. \ref{10-scalarcharge-M} shows that the mass at which the scalar charge again becomes zero reduces as $\beta$ increases. In addition, the maximum value of scalar charge decreases as the coupling constant grows.
The stars with nonzero scalar charge are the scalarized neutron stars. The range of mass with nonzero scalar charge is the same in two models. However, M2 predicts more scalar charge compared to M1. Because the neutron stars are more scalarized in M2.
\begin{figure}[h]
	{\includegraphics[scale=0.67]{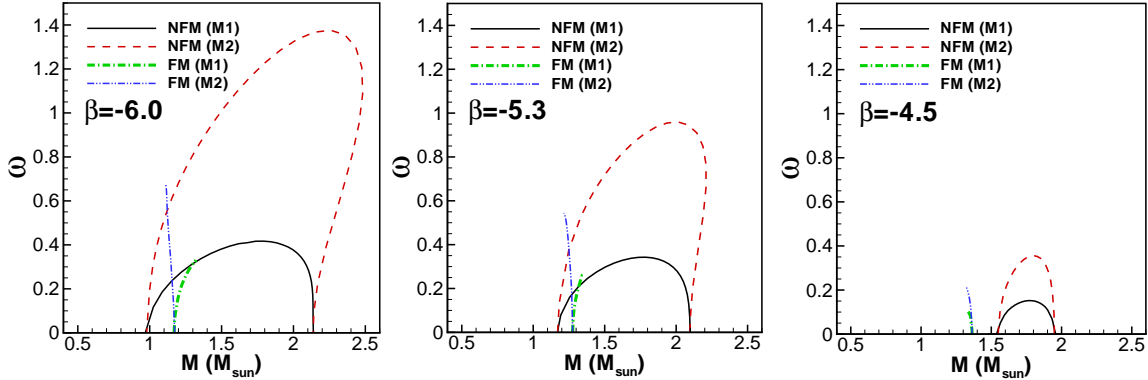}
\label{10-scalarcharge-M}	}
		\caption{Scalar charge, $\omega$, versus the mass, $M$, for
		 NFM and FM neutron stars applying M1 and M2 with different values of the coupling constant, $\beta$.}
	\label{10-scalarcharge-M}
\end{figure}
For FM stars in M1 with lower values of $\beta$, i.e. $-6.0$ and $-5.3$, the scalar charge is an increasing function of mass. This is due to the fact that in these conditions, the scalar field is an increasing function of both density and mass (see Figs. \ref{2-m-rho} and \ref{6-phi-rho}). However, for FM stars in M1 with $\beta=-4.5$ and also in M2 with different values of $\beta$, the scalar charge decreases when the mass grows. It is due to the fact that according to Figs. \ref{2-m-rho} and \ref{6-phi-rho}, by increasing the density the scalar field increases while the mass decreases.
Therefore, for FM stars, the slope of scalar charge versus the mass depends on the model of coupling function as well as the coupling constant. By increasing $\beta$, the maximum value of scalar charge in FM neutron stars like the NFM ones decreases. The maximum value of scalar charge in NFM stars in two models with different values of
$\beta$ is greater than the one in FM stars. The curve related to the scalar charge of FM neutron stars
is not a closed curve unlike the one related to the NFM stars. This is a result of this fact that the FM neutron stars remain scalarized up to high densities.

\begin{figure}[h]
	{\includegraphics[scale=0.67]{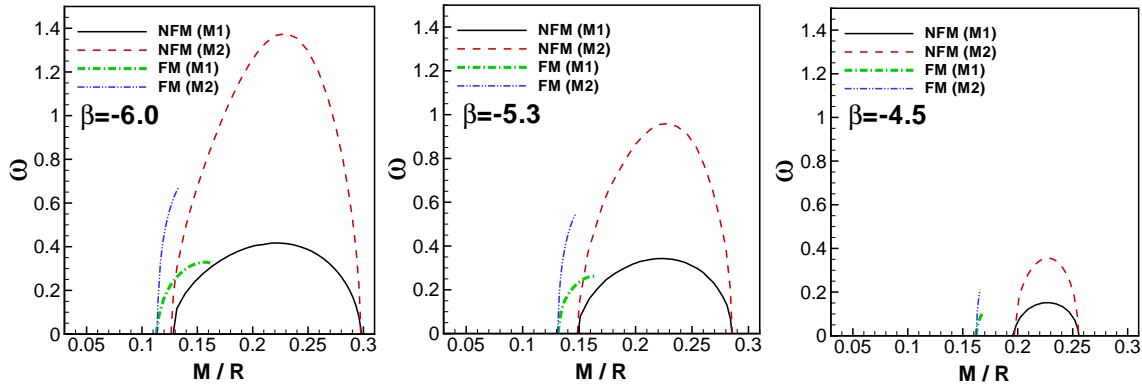}
	}
		\caption{Same as Fig. \ref{10-scalarcharge-M} but for the scalar charge versus the compactness of star, $M/R$.}
	\label{11-scalarcharge-CMP}
\end{figure}

Fig. \ref{11-scalarcharge-CMP} presents the scalar charge of NFM and FM neutron stars versus their compactness. In two models with all values of $\beta$ for NFM and FM neutron stars with low compactness, the scalar charge is zero.
In NFM neutron stars with the compactness higher than a special value, the scalar charge increases with the compactness. Besides, the scalar charge again becomes zero for NFM stars with high compactness. However, for FM stars, the scalar charge increases monotonically with the increase of the compactness. Fig. \ref{11-scalarcharge-CMP} confirms that the compactness at which the scalar charge becomes nonzero is higher for NFM neutron stars compared to FM ones.
This means that the FM stars with lower compactness can also have scalar charge. For all stars in two models, the value
of compactness at which the scalar charge becomes nonzero increases when the coupling constant grows. It is due to the fact that with lower values of $|\beta|$ and approaching to GR, the more compactness is needed to have the scalar charge.
In addition, for both NFM and FM neutron stars in two models, the range of compactness at which the scalar charge is nonzero decreases by increasing $\beta$. It means that with the lower values of $|\beta|$ and approaching to GR, the chance for finding the stars with scalar charge is lower.
\section{Summary and Concluding Remarks}

The structure of ferromagnetic (FM) neutron stars in scalar-tensor theories of gravity has bee studied.
To describe the neutron star, we employ the equation of state of FM
neutron matter with Skyrme-type interactions. We found that the soft
EOS of FM neutron matter leads to the lower values of the neutron star mass compared to the non-ferromagnetic (NFM) one.
Our results confirm that with the lower values of the coupling constant, the results of STT deviate significantly from the ones in GR. In the cases which the results of STT are different from GR, the neutron stars are scalarized.
For high density NFM neutron stars, the results of STT and GR are the same and there is no
scalarized NFM neutron stars with high central density. However, for the FM
neutron stars up to high central density considered in this work, the results of GR and STT are not the same and these stars
even with high densities are scalarized. We found that the densities at which the scalarization takes place are not equal for NFM and FM neutron stars. Our calculations show that for both NFM and FM neutron stars, the first critical density of scalarization increases as the coupling constant grows.
In addition, in NFM stars, the second critical density decreases as the coupling constant increases. The range of scalarization in FM neutron stars is greater than NFM ones. For both NFM and FM stars, the maximum value of the central scalar field reduces as the coupling constant increases. We showed that the deviation of STT from GR in the mass-radius relation is more significant with lower values of the coupling constant. Besides, this deviation in the mass-radius relation as well as the
scalarization are seen in a greater region when the coupling constant takes lower values. We found that the deviation of STT from GR is more significant for NFM stars compared to FM ones. It means that the EOS of neutron
matter affects the amount of deviation of STT from GR. For NFM neutron
stars, the scalarized ones are larger in size compared to the GR solutions. Moreover, for FM neutron
stars, the mass of scalarized stars is lower than the ones in GR.
In FM neutron stars, the maximum value of the coupling constant at which the stars are scalarized
is higher compared to the NFM neutron stars.
Our results verify that
the compactness at which the scalar charge becomes nonzero is greater for NFM neutron stars compared to FM ones. Our work determines the magnetic effects of neutron stars on the
properties of these stars in the STTs, i.e.
the profile of scalar field, the scalarization and its critical densities, the scalar charge, and
the deviation of STT from GR. In fact, we conclude that when one considers the neutron stars in ferromagnetic phase
within the STTs, it is necessary to note that the neutron star EOS has significant effects on the behaviour of these
stars in STTs.
Indeed, to test the scalar-tensor theories of gravity by neutron stars which are one of the best laboratories
for high energy physics, because of the magnetic properties of these compact objects, it is more proper to
note our results when one tests the scalar-tensor theories of gravity.
\acknowledgements{We would like to
thank Shiraz University Research Council.}



\end{document}